\documentclass[%
 aip,
 amsmath,amssymb,
 reprint,%
]{revtex4-1}
\usepackage{xcolor}
\usepackage{graphicx}
\usepackage{dcolumn}
\usepackage{bm}
\usepackage[utf8]{inputenc}
\usepackage[T1]{fontenc}
\usepackage{mathptmx}
\usepackage{etoolbox}

\makeatletter
\def\@email#1#2{%
 \endgroup
 \patchcmd{\titleblock@produce}
  {\frontmatter@RRAPformat}
  {\frontmatter@RRAPformat{\produce@RRAP{*#1\href{mailto:#2}{#2}}}\frontmatter@RRAPformat}
  {}{}
}%
\makeatother
\begin{document}

\preprint{AIP/123-QED}

\title[Magnetic Materials for Quantum Magnonics]{Magnetic Materials for Quantum Magnonics}
\author{Rostyslav O. Serha}
 \altaffiliation[Author to whom correspondence should be addressed: ]{\\rostyslav.serha@univie.ac.at}
 \affiliation{Faculty of Physics, University of Vienna, 1090 Vienna, Austria.}
 \affiliation{Vienna Doctoral School in Physics, University of Vienna, 1090 Vienna, Austria.}
 
\author{Carsten Dubs}%
\affiliation{ 
INNOVENT e.V. Technologieentwicklung, 07745 Jena, Germany.
}%

\author{Andrii V. Chumak}
\affiliation{%
Faculty of Physics, University of Vienna, 1090 Vienna, Austria.
}%

\date{\today}

\begin{abstract}
Quantum magnonics studies the quantum properties of magnons\,---\,the quanta of spin waves\,---\,and their application in quantum information processing. Progress in this field depends on identifying magnetic materials with characteristics tailored to the diverse requirements of magnonics and quantum magnonics. For single-magnon excitation, its control, hybrid coupling, and entanglement, the most critical property is the ability to support long magnon lifetimes. This perspective reviews established and emerging magnetic materials\,---\,including ferromagnetic metals, Heusler compounds, antiferromagnets, altermagnets, organic and 2D van der Waals magnets, hexaferrites, europium chalcogenides, and in particular yttrium iron garnet (YIG)\,---\,highlighting their key characteristics. YIG remains the benchmark, with bulk crystals supporting sub-microsecond Kittel-mode lifetimes and ultra-pure spheres achieving $\sim$18\,µs for dipolar-exchange magnons at millikelvin temperatures. However, thin YIG films on gadolinium gallium garnet (GGG) substrates suffer from severe lifetime reduction due to substrate-induced losses. In contrast, YIG films on a new lattice matched, diamagnetic alternative, yttrium scandium gallium/aluminum garnet (YSGAG), overcomes these limitations and preserves low magnetic damping down to millikelvin temperatures. These advances provide a practical pathway toward ultralong-living magnons in thin films, enabling scalable quantum magnonics with coherent transport, strong magnon-photon, magnon-qubit coupling, and integrated quantum networks.

\end{abstract}

\maketitle

\section{\label{sec:Intro} Introduction into quantum magnonics}

Quantum computing promises transformative advances in computation, with potential impacts on cryptography, materials science, and artificial intelligence~\cite{Nielsen2010,Arute2019}. A central challenge for solid-state quantum technology is to identify an on-chip platform with a suitable nanoscale information carrier~\cite{Chumak2022}. Magnons\,---\,the bosonic quanta of spin-wave excitations in ordered magnets\,---\,are promising for this role. The emerging field of quantum magnonics~\cite{Lachance-Quirion2019,Li2020,Li2022,Chumak2022,Yuan2022,Zhang2023} operates with quantum states of magnons, including single magnons, typically at millikelvin temperatures, where thermal magnons are suppressed. Magnons cover gigahertz to terahertz frequencies~\cite{Levchenko2024}, scale to nanometer footprints~\cite{Wang2019b,Heinz2020}, interface naturally with microwave circuitry~\cite{Chumak2017}, and exhibit strongly nonlinear and nonreciprocal dynamics~\cite{Chumak2014,Wang2024,Pirro2021,QWang2023,Zenbaa2025}. These properties position magnons as promising candidates for quantum information processing, coherent transport, and entangled-state preparation~\cite{Lachance-Quirion2019,Li2020,Li2022,Chumak2022,Yuan2022}.

At room temperature (RT), macroscopic quantum phenomena in magnon gases are already accessible. Theory and experiment have established Bose--Einstein condensation of magnons \cite{Demokritov2006,Serga2014,Schneider2020,Schneider2021}, magnonic supercurrents \cite{Bozhko2016,Kreil2018}, and the magnonic Aharonov--Casher effect \cite{Serha2023,Nakata2014}. In antiferromagnets, quantum magnonics can reach the terahertz regime. Cavity-coupled antiferromagnetic magnons have been demonstrated in collinear and canted phases \cite{Boventer2023}, and terahertz cavity magnon polaritons have been realized in prototype platforms \cite{Kritzell2023}. Long-distance coherent antiferromagnetic spin-wave propagation at RT further supports device scaling viability \cite{HWang2023}. These advances complement broader arguments for magnons as quantum excitations with frequency tunability and compatibility with hybrid architectures \cite{Zhang2023,Yuan2022,Chumak2022,Li2020}.\looseness=-3

Single-magnon operations in ferromagnetically ordered media require suppression of thermal magnon populations. At a frequency of 4\,GHz, which is typical for superconducting qubits and sufficiently high to forbid three-magnon splitting while providing longer lifetimes than at higher frequencies where viscous damping is stronger~\cite{Gurevich1996}, the thermal magnon occupation can be kept below 0.01 according to Bose--Einstein statistics only if experiments are performed in dilution refrigerators at temperatures of about 40\,mK or lower~\cite{Clerk2010}. The objective is to generate, control, and read out single magnons in hybrid systems that couple to photons, microwave cavities, phonons, or superconducting qubits \cite{Tabuchi2015,Wolski2020,Lachance-Quirion2020,Baity2021,Zuo2024}.\looseness=-3

Theory identifies roles for magnons as quantum interconnects and as probes of solid state quantum effects \cite{Kounalakis2022,Jiang2023, Dols2024, Ballestero2025}. Further proposals predict steady state magnon squeezing and photon-magnon entanglement in cavity hybrids using Kerr nonlinearities and engineered couplings \cite{Kostylev2019,Yang2021,Haghshenasfard2020,Toklikishvili2023, Sharma2024}. Magnomechanical schemes show how squeezing can enhance correlations and light-matter control and enable nonclassical state preparation \cite{Lu2023,Zhang2024}. Experimentally, key milestones include coherent coupling of fundamental magnon modes to superconducting resonators \cite{Lachance-Quirion2019,Li2022,Song2025}, qubit based quantum sensing \cite{Wolski2020}, single magnon detection with a superconducting qubit \cite{Lachance-Quirion2020}, full Wigner function tomography of a single magnon \cite{Xu2023}, macroscopic Bell state between a spin system and a superconducting qubit\cite{Xu2024}, and NV-based platforms that quantify magnon mediated NV-NV coupling \cite{Fukami2024,Borst2023}. Although the best Kittel-mode magnon lifetimes \cite{Lachance-Quirion2020,Xu2023} of several hundred nanoseconds are shorter than the decoherence times of state-of-the-art superconducting qubits, which reach tens of microseconds\cite{Krantz2019}, they are sufficient for entanglement transfer and for interfacing heterogeneous quantum systems \cite{Tabuchi2015,Li2020,Lachance-Quirion2019,Yuan2022,Zhang2023}. Recently, a qubit-free approach to quantum magnonics has been proposed, where Gaussian quantum states of ferromagnetic resonance magnons, including squeezed\cite{Haghshenasfard2016} and entangled states, are generated via parametric driving and magnonic nonlinearities, enabling quantum-state preparation and verification using conventional ferromagnetic resonance techniques\cite{Mycroft2026}.

The most recent single-magnon experiments have been carried out on bulk spheres of yttrium iron garnet (YIG) \cite{Lachance-Quirion2020,Xu2023,Xu2024}, which provides the longest coherence with Kittel-mode lifetimes generally below one microsecond \cite{YIGmagnonics}, underscoring its importance as the leading platform for quantum magnonics. Significant progress has been made in extending lifetimes at low temperatures, at which in ultra-pure bulk YIG, dipolar-exchange magnons reach about 18\,µs at millikelvin temperatures \cite{Serha2025b}.

A central challenge in integrating superconducting qubits with magnetic resonant systems is the strong sensitivity of conventional qubits to external magnetic fields, which significantly reduces their coherence times. Such fields are, however, required to tune magnetic resonators to microwave frequencies relevant for quantum applications. For instance, the Kittel mode of a YIG sphere at 4\,GHz requires an external magnetic field of approximately 150\,mT, a field strength that typical aluminum-based superconducting qubits cannot tolerate. Current experimental approaches therefore rely on spatially separating the magnetic resonator and the superconducting qubit, coupling them indirectly via a microwave cavity mode. This allows the YIG sphere to be biased by a magnetic field while keeping the qubit in a field-free environment\cite{Lachance-Quirion2020,Xu2023,Xu2024}. This strategy could be further improved by employing planar superconducting resonators fabricated from materials that are more tolerant to magnetic fields such as niobium, niobium–tin or niobium-titanium, or by using qubits intrinsically robust against moderate magnetic fields\cite{PitaVidal2020,Kringhj2021,Gnzler2025}. An alternative route is the use of magnetic waveguides as magnonic media operating at or near zero external field. YIG waveguides have already demonstrated operation at frequencies around 1.7\,GHz\cite{Nikolaev2023}. Upon cooling to cryogenic temperatures, the increase in saturation magnetization\cite{Serha2025c} is expected to shift this frequency upward to approximately 3\,GHz, placing it within a range that is well compatible with commonly engineered superconducting qubits.

Scaling from localized resonators to circuit-level links requires propagating magnons. Propagation enables spatial separation of single-magnon sources and detectors, and allows the transport of quantum states across a chip \cite{Karenovska2018,karenowska2015excitation,Knauer2023}. This shifts attention from bulk magnetic resonators to thin films, where dipolar-driven magnons can travel. However, thin films perform worse than bulk samples, as their higher surface-to-volume ratio  enhances the effect of surface defects. A further challenge in thin YIG films arises from the paramagnetism of GGG substrates at low temperatures \cite{Serha2025,Serha2024,Guo2022,Jermain2017,Michalceanu2018,Kosen2019}. This limitation has recently been overcome by growing YIG films on a newly developed lattice-matched diamagnetic substrate \cite{Dubs2025,Serha2025d}. These advances enable on-chip platforms that excite and detect propagating magnons with planar resonators \cite{Wang2024}. They also connect to ultralow-temperature studies of damping in bulk YIG \cite{Serha2025b,Kosen2019}, to wavenumber-dependent damping in films \cite{Schmoll2024}, and to nanoscale propagation at cryogenic temperatures \cite{Knauer2023}. Together, this progress establishes a foundation for quantum magnonic circuits that integrate superconducting qubits, photons, and phonons, and exploit the nonlinear and nonreciprocal dynamics of magnons \cite{Barman2021,Flebus2024,Chumak2015,Chumak2022}.\vspace{-0.0em}

\section{\label{sec:Mat} Available magnetic materials}
\begin{figure*}
\includegraphics{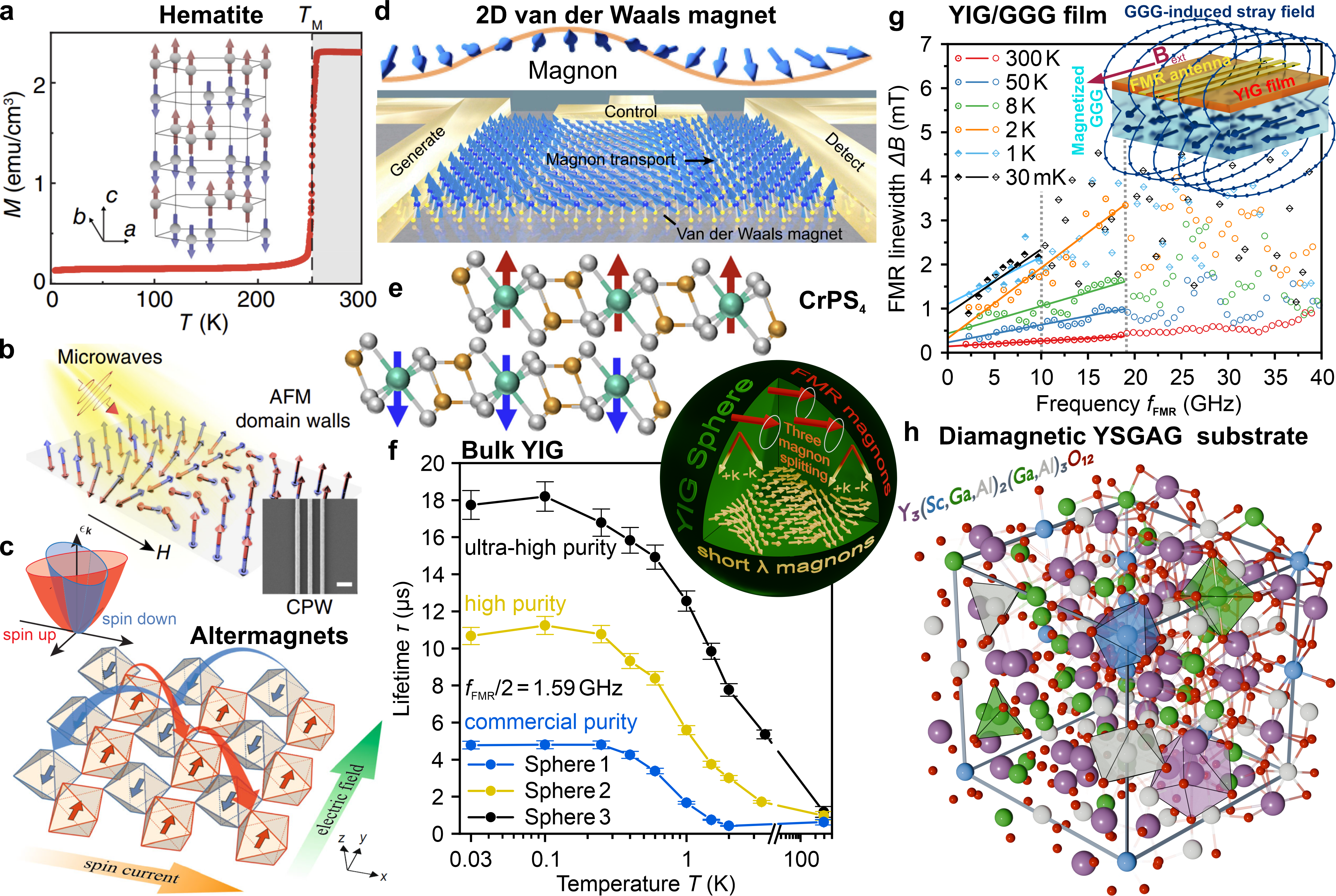}
\caption{(a) Temperature-dependent magnetization of hematite. Inset shows the easy-axis crystal structure for \(T<T_{\mathrm{M}}\)\cite{Chen2025}. Reproduced with permission from J.~Chen \textit{et al.}, \textit{Phys. Rev. Lett.} \textbf{134}, 056701 (2025). Copyright 2025 American Physical Society.
(b) Sketch of the antiferromagnetic (AFM) spin texture under microwave excitation. The magnetic field is applied along the \(x\) direction. Inset shows a SEM image of the coplanar waveguide (CPW); scale bar \(500~\mathrm{nm}\)\cite{Chen2025}. Reproduced with permission from J.~Chen \textit{et al.}, \textit{Phys. Rev. Lett.} \textbf{134}, 056701 (2025). Copyright 2025 American Physical Society.
(c) Altermagnet: d-wave–type spin-split bands (upper left) and perovskite schematic with sublattice-dependent, anisotropic spin-dependent hoppings (red/blue arrows)\cite{Naka2025}. Reproduced from M.~Naka \textit{et al.}, \textit{npj Spintronics} \textbf{3}, 1 (2025), licensed under a Creative Commons Attribution--NonCommercial--NoDerivatives 4.0 International License.
(d) Illustration of spin waves in a 2D van der Waals magnet. Magnon spintronics relies on manipulation and control of magnon spin transport from an injector to a detector. The electrical contacts for generating, controlling, and detecting magnons are shown as gold pads, while the van der Waals magnet is depicted by blue and yellow atoms (magnetic and non-magnetic, respectively). The spin is represented by a blue arrow whose orientation varies spatially to indicate the transported magnon\cite{MaasValero2025}. Reproduced from S.~Mañas-Valero \textit{et al.}, \textit{Fundamentals and applications of van der Waals magnets in magnon spintronics}, \textit{Newton} \textbf{1}, 100018 (2025). Copyright Elsevier (2025).
(e) Atoms and spins in a bilayer of the van der Waals antiferromagnet CrPS\(_4\). Red and blue arrows indicate the local magnetic moments of the Cr atoms (turquoise). The interlayer (intralayer) exchange coupling is ferromagnetic (antiferromagnetic)\cite{deWal2023}. Reproduced with permission from D.~K.~de~Wal \textit{et al.}, \textit{Phys. Rev. B} \textbf{107}, L180403 (2023). Copyright 2023 American Physical Society.
(f) Lifetime \(\tau\) of secondary dipolar–exchange magnons (DESW) at \(f_{\mathrm{FMR}}/2\) as a function of temperature \(T\) (logarithmic \(T\)-axis) for three YIG spheres. The purity of the YIG samples with respect to rare-earth ion impurities increases from commercial standard purity in Sphere 1 to ultra-high purity in Sphere 3. Illustration shows a schematic of three-magnon splitting\,—\,the \(k=0\) uniform mode (\(f_{\mathrm{FMR}}=3.17~\mathrm{GHz}\)) splits into two counter-propagating magnons at \(1.59~\mathrm{GHz}\) with $|k|\approx 3~\mathrm{rad}\,$µ$\mathrm{m}^{-1}$ in the dipolar--exchange spin-wave regime\cite{Serha2025b}. Reproduced from R.~O.~Serha \textit{et al.}, arXiv:2505.22773 (2025); licensed under a Creative Commons Attribution 4.0 International (CC BY) license.
(g) Experimental FMR linewidth \(\Delta f\) as a function of the FMR frequency \(f_{\mathrm{FMR}}\) for a \(100~\mathrm{nm}\) YIG film on a GGG substrate. The half-solid points are measurement points up to \(18~\mathrm{GHz}\) for temperatures above \(2~\mathrm{K}\) and up to \(10~\mathrm{GHz}\) for temperatures below \(2~\mathrm{K}\) (gray dashed lines mark these limits). The straight lines are Gilbert fits performed up to \(18~\mathrm{GHz}\) and \(10~\mathrm{GHz}\), respectively. Above these values, the linewidth shows non-Gilbert behavior versus \(f_{\mathrm{FMR}}\). The linewidth broadening and deviation from Gilbert behavior are due to magnetization of the GGG substrate at low temperatures (inset), which creates an inhomogeneous stray field affecting the internal field of the YIG film\cite{Serha2025}. Reproduced from R.~O.~Serha \textit{et al.}, \textit{Materials Today Quantum} \textbf{5}, 100025 (2025); licensed under a Creative Commons Attribution 4.0 International (CC BY) license.
(h) Crystal structure of diamagnetic YSGAG, an alternative substrate to paramagnetic GGG for YIG. The cubic lattice structure of YSGAG is similar to that of YIG, but with iron ions replaced at the octahedral sites by scandium, gallium, and aluminum, while tetrahedral sites are occupied by gallium and aluminum\cite{Serha2025d}. Reproduced from R.~O.~Serha \textit{et al.},  arXiv:2508.19044 (2025); licensed under a Creative Commons Attribution-NonCommercial-NoDerivatives 4.0 International (CC BY-NC-ND) license.
}\label{f:1}
\end{figure*}

Magnonics relies on ordered magnetic media that efficiently support spin waves. In this Perspective, we briefly present and analyze different material platforms, including those that have been widely used for many years as well as more recent and highly attractive candidates. An overview is provided in Tab.\,\ref{tab:materials}, and representative examples are illustrated in Fig.\,\ref{f:1}. Further, the magnon lifetime in the discussed materials is commonly extracted from resonance measurements of the fundamental Kittel mode. The full width at half maximum (FWHM) $\Delta B$ of the magnetic resonance quantifies the magnetic losses of the system and allows one to determine the lifetime $\tau$ of the magnetic excitation via
\begin{equation}
    \tau = \frac{1}{\gamma \cdot \Delta B}\,,
\end{equation}
where $\gamma$ is the gyromagnetic ratio\cite{Gurevich1996}. 
When comparing magnon damping and lifetimes across different magnetic materials, it is important to distinguish between the phenomenological Gilbert damping parameter $\alpha$ and the actual lifetime $\tau$ of a magnonic excitation. While $\alpha$ provides a convenient description of viscous damping in the Landau--Lifshitz--Gilbert framework and is commonly extracted from resonance linewidths, it does not uniquely determine magnon lifetimes in general. The magnetic linewidth $\Delta B$ of a system is related to the Gilbert damping parameter $\alpha$ by
\begin{equation}
\Delta B(f)=\Delta B_0+\frac{4\pi\alpha}{\gamma}\,f\,,
\end{equation}
where $\Delta B_0$ denotes the inhomogeneous linewidth broadening, corresponding to the linearly extrapolated linewidth at zero frequency ($f=0$).

In practice, linewidth broadening and magnon relaxation can arise from a variety of mechanisms beyond intrinsic viscous Gilbert damping, including impurity scattering, two-magnon scattering, magnon--phonon coupling, and slow-relaxor processes on rare-earth and $L$-state transition metal ion impurities, the latter being particularly relevant in YIG at cryogenic temperatures\cite{Gurevich1996,PardaviHorvath2000}. As a result, linewidth measurements may deviate substantially from those expected from a simple linear Gilbert damping picture, especially when comparing different magnon wavelengths, temperature regimes, or material classes.

\newcommand{\pcell}[2]{\parbox[c]{#1}{\centering #2}}
\newcommand{\pcellL}[2]{\parbox[c]{#1}{\raggedright #2}}
\begin{table*}[t!]
\caption{\label{tab:materials}Comparison of magnetic media relevant to quantum magnonics.}
\begingroup
\renewcommand{\arraystretch}{1.3}
\setlength{\tabcolsep}{6pt}
\begin{ruledtabular}
\begin{tabular}{p{0.145\textwidth} p{0.093\textwidth} p{0.093\textwidth}
                p{0.093\textwidth} p{0.093\textwidth} p{0.093\textwidth} p{0.12\textwidth}p{0.093\textwidth}}
\textbf{Property} &
\pcell{0.093\textwidth}{YIG\\\cite{SagaYIG,Serha2025c,Bertaut1956,YIGmagnonics,Serha2025b,Geller1957,Klingler2017,Klingler2015,Maier-Flaig2017}} &

\pcell{0.093\textwidth}{Permalloy (Py)\cite{Demidov2015, Kalarickal2006, Sebastian2015b, Patton1968}} &
\pcell{0.093\textwidth}{CoFeB\\\cite{Brunsch1979,Liu2011,Conca2013,Zenbaa2025}} &
\pcell{0.093\textwidth}{Hematite\\\cite{Chen2025,Hamdi2023,HWang2023,Samuelsen1970}} &
\pcell{0.093\textwidth}{Hexaferrite BaM \cite{song2010self, Harris2012, malkinski2012advanced, zhu2021mechanical, popov2022plane}} &
\pcell{0.12\textwidth}{Heusler CMFS\cite{Liu2009,Trudel2010,Oogane2010,Sebastian2012}} &
\pcell{0.093\textwidth}{CrPS\\\cite{deWal2023,Budko2021,Lee2017,Freeman2025}}\\[5pt]
\hline\\[-10pt]
\pcellL{0.155\textwidth}{Chemical composition} &
\pcell{0.093\textwidth}{$\mathrm{Y_3Fe_5O_{12}}$} &

\pcell{0.093\textwidth}{$\mathrm{Ni_{81}Fe_{19}}$} &
\pcell{0.093\textwidth}{$\mathrm{Co_{40}Fe_{40}B_{20}}$} &
\pcell{0.093\textwidth}{$\mathrm{Fe_{2}O_{3}}$} &
\pcell{0.093\textwidth}{$\mathrm{BaFe_{12}O_{19}}$} &
\pcell{0.12\textwidth}{$\mathrm{Co_2Mn_{0.6}Fe_{0.4}Si}$}&
\pcell{0.093\textwidth}{$\mathrm{CrPS_4}$}\\[8pt]
\pcellL{0.155\textwidth}{Structure} &
\pcell{0.093\textwidth}{single-crystalline} &

\pcell{0.093\textwidth}{polycrystalline} &
\pcell{0.093\textwidth}{amorphous} &
\pcell{0.093\textwidth}{single-crystalline} &
\pcell{0.093\textwidth}{single-crystalline} &
\pcell{0.12\textwidth}{single-crystalline} &
\pcell{0.093\textwidth}{2D van der Waals}\\[8pt]
\pcellL{0.155\textwidth}{Gilbert damping $\alpha$} &
\pcell{0.093\textwidth}{$3\times10^{-5}$} &

\pcell{0.093\textwidth}{$7\times10^{-3}$} &
\pcell{0.093\textwidth}{$4\times10^{-3}$} &
\pcell{0.093\textwidth}{$5\times10^{-5}$} &
\pcell{0.093\textwidth}{$7\times10^{-4}$} &
\pcell{0.12\textwidth}{$3\times10^{-3}$}&
\pcell{0.093\textwidth}{$1\times10^{-2}$ \footnotemark[1]}\\[8pt]
\pcellL{0.155\textwidth}{Saturation~magnetization @ RT $M_\mathrm{_S}$\,(kA/m)} &
\pcell{0.093\textwidth}{140} &

\pcell{0.093\textwidth}{800} &
\pcell{0.093\textwidth}{1250} &
\pcell{0.093\textwidth}{---} &
\pcell{0.093\textwidth}{320-390} &
\pcell{0.12\textwidth}{1000}&
\pcell{0.093\textwidth}{---}\\[8pt]
\pcellL{0.155\textwidth}{Exchange constant \\@ RT $A$ (pJ/m)} &
\pcell{0.093\textwidth}{3.6} &

\pcell{0.093\textwidth}{16} &
\pcell{0.093\textwidth}{15} &
\pcell{0.093\textwidth}{$10-20$} &
\pcell{0.093\textwidth}{6-9} &
\pcell{0.12\textwidth}{13}&
\pcell{0.093\textwidth}{10 \footnotemark[1]} \\[8pt]
\pcellL{0.155\textwidth}{Curie~(Néel)~temperature $T_{\mathrm{_{C(N)}}}$ (K)} &
\pcell{0.093\textwidth}{560} &

\pcell{0.093\textwidth}{$550-870$} &
\pcell{0.093\textwidth}{1000} &
\pcell{0.093\textwidth}{950} &
\pcell{0.093\textwidth}{726} &
\pcell{0.12\textwidth}{$>985$}&
\pcell{0.093\textwidth}{37}\\

\end{tabular}
\footnotetext[1]{Used in micromagnetic simulations\cite{Freeman2025}.}
\end{ruledtabular}
\endgroup
\end{table*}

In antiferromagnets, the role of Gilbert damping and its relation to magnon lifetimes differ significantly from those in ferro- and ferrimagnets due to the two-sublattice nature of the magnetic order\cite{Gomonay2018}. The dynamics are governed by the Néel order parameter and described by inertial equations of motion, with damping entering as a matrix rather than as a single scalar parameter\cite{Kamra2018,Simensen2020}. Consequently, direct comparisons of Gilbert damping parameters across different magnetic orders are generally not meaningful without accounting for the underlying relaxation mechanisms. Therefore, in this Perspective we focus on direct linewidth measurements of magnetic excitations in ferro- and ferrimagnets as a more reliable measure of the actual magnon lifetime.

\subsection*{Ferromagnetic metals}
A common class of materials used to excite spin waves is ferromagnetic metals, valued for their straightforward fabrication, compatibility with standard nanofabrication techniques, and high saturation magnetization, which enables efficient spin-wave excitation and access to large frequency bandwidths (see Tab.\,\ref{tab:materials}). Prominent examples include Permalloy ($\mathrm{Ni_{81}Fe_{19}}$)\cite{Demidov2015,Kalarickal2006,Patton1968}, CoFeB ($\mathrm{Co_{40}Fe_{40}B_{20}}$)\cite{Talmelli2020,Wojewoda2024,Brunsch1979,Liu2011,Conca2013,Zenbaa2025}, and CoFe\cite{Schoen2016}. In particular, propagating spin-wave spectroscopy on epitaxial single-crystal Fe thin films revealed an effective Gilbert damping as low as $\alpha \approx 2.5\times10^{-3}$, with lifetimes of up to 2\,ns\cite{Gladii2017}. Even lower intrinsic damping has been demonstrated in carefully engineered CoFe alloys, where Gilbert damping values down to $\alpha \approx 5\times10^{-4}$ were reported\cite{Schoen2016}. This reduction is attributed to a minimum in the electronic density of states at the Fermi level, which suppresses magnon–electron scattering and highlights the role of band-structure engineering in controlling dissipation in metallic systems. Their relatively large exchange stiffness and saturation magnetization further allow for short-wavelength magnons and high group velocities, advantageous for compact device architectures and broadband operation\cite{Levchenko2024}.

At the same time, the presence of itinerant electrons in ferromagnetic metals introduces intrinsic dissipation channels through electron–magnon and electron–phonon scattering, resulting in magnetic damping that typically limits magnon lifetimes to the nanosecond regime\cite{Korenman1972}. Despite short lifetimes ferromagnetic metals remain relevant for cryogenic magnonics and hybrid quantum platforms, particularly in regimes requiring fast dynamics, large saturation magnetization, electrical tunability, strong nonlinearities, or strong coupling to microwave photons\cite{Li2019}. Their metallic nature enables efficient electrical control via spin-transfer and spin–orbit torques and supports rapid energy exchange in strongly coupled hybrid systems, facilitating the integration of magnonic and electronic functionalities on a single chip\cite{Zhang2023}.

\subsection*{Heusler compounds}

Heusler compounds, a class of mostly ternary or quaternary intermetallics with highly tunable electronic and magnetic properties\cite{Graf2011,Elphick2021,Mantion2024}, offer an improvement over conventional metallic ferromagnets by combining high spin polarization, large saturation magnetization, and comparatively low magnetic damping. Among the most studied in the magnonics and spintronics communities are Co$_2$-based full Heusler alloys, where half-metallic electronic structures and high crystalline order can suppress magnon dissipation channels\cite{Guillemard2019}.

A particularly promising example is the quaternary Heusler compound CMFS ($\mathrm{Co_2Mn_{0.6}Fe_{0.4}Si}$), for which low-damping spin-wave propagation and extended lifetimes have been demonstrated experimentally\cite{Sebastian2012,ChumakHandbook2019}. Thin film studies of CFMS and related Co$_2$FeGa$_{0.5}$Ge$_{0.5}$ (CFGG) alloys report effective Gilbert damping parameters in the range of $\alpha_{\rm eff}\sim10^{-3}$ to $10^{-2}$, with evidence for both intrinsic and extrinsic damping contributions, such as two-magnon scattering\cite{OChumak2021}. Additionally, substrate engineering has been shown to reduce Gilbert damping in CFAS (Co$_2$FeAl$_{0.5}$Si$_{0.5}$) Heusler films, with optimal annealing conditions yielding $\alpha$ values significantly lower than in as-grown films\cite{Love2021}.

Furthermore, theoretical and experimental studies on Co$_2$MnZ (Z = Si, Ge, Sn, Sb) Heusler systems have identified pathways toward ultralow magnetic damping ($\alpha \sim 4 \times 10^{-4}$–$9 \times 10^{-4}$), representing some of the lowest values reported for metallic ferromagnets and highlighting the crucial role of spin polarization and Fermi-level band structure in damping reduction\cite{Guillemard2019}. More recently, ultra-low magnetic damping has been demonstrated in epitaxial Co$_2$MnSi Heusler thin films, where transport measurements revealed a strong correlation between crystalline order, spin polarization, and reduced magnon dissipation, underscoring the potential of highly ordered Heusler alloys for low-loss magnon transport down to cryogenic temperatures\cite{deMelo2021}. In addition, other Co-based Heusler compounds, such as CoFe$_2$Al, have also been reported to exhibit relatively low magnetic damping, suggesting a broader materials landscape for low-loss metallic magnonics\cite{RWang2023}.

Despite these improvements relative to conventional ferromagnetic metals, Heusler compounds still possess finite electrical conductivity, which continues to provide intrinsic dissipation channels via electron–magnon and electron–phonon scattering. Consequently, magnon lifetimes in optimized Heusler systems remain in the nanosecond regime. As a result, Heusler compounds represent an attractive and versatile materials platform for magnonics and spintronics, particularly in hybrid device architectures and electrically controlled magnon transport, while ongoing advances in materials optimization continue to expand their potential for long-lived quantum magnon applications.

\subsection*{Antiferromagnets}
Antiferromagnets have emerged as a promising platform for magnonics and quantum magnonics, driven by a series of recent theoretical advances. Foundational work has shown that antiferromagnetic insulators can support superfluid spin transport, enabling coherent long-range spin currents without net magnetization \cite{Takei2014}. Their rich spin textures and ultrafast dynamics further highlight the potential of antiferromagnets for spintronic and magnonic devices \cite{Gomonay2018}. More recently, the quantum aspects of antiferromagnetic magnons have been explored. In this context, a transmon qubit has been proposed as a probe of their quantum characteristics\cite{AzimiMousolou2023}. In parallel, spin–orbit interaction has been predicted to mediate the hybridization of antiferromagnetic magnons with plasmons, i.e., collective charge-density oscillations of conduction electrons, resulting in coherently coupled magnon–plasmon modes with modified dispersions and nontrivial band topology described within a quantum-mechanical framework\cite{Dyrda2023}.

A representative antiferromagnet of growing interest in magnonics is hematite\cite{Gckelhorn2022,ElKanj2023,Gckelhorn2023} ($\alpha$-Fe$_2$O$_3$) and shown in Fig.\,\ref{f:1}\,(a). Bulk studies at GHz frequencies report low Gilbert damping (see Tab.\,\ref{tab:materials}) \cite{Chen2025,Hamdi2023,HWang2023}. In the easy-axis phase below the Morin transition, all-electrical spectroscopy reveals nearly gapless antiferromagnetic magnons bound to $180^\circ$ domain walls that propagate over micrometre distances with a group velocity of about $6\,\mathrm{km/s}$, as illustrated in Fig.\,\ref{f:1}\,(b). The domain-wall mode exhibits a full width at half maximum (FWHM) frequency linewidth $\Delta f_{\mathrm{FWHM}} \approx 1.2\,\mathrm{GHz}$ and a corresponding lifetime $\tau \approx 0.3\,\mathrm{ns}$ \cite{Chen2025}. Another canted antiferromagnet with even better damping characteristics is yttrium orthoferrite ($\mathrm{YFeO_3}$) \cite{Das2022}. It exhibits linewidths below 10\,mT at frequencies around 350\,GHz and a temperature of 20\,K, corresponding to magnon lifetimes slightly above one nanosecond. Although antiferromagnets can reach terahertz frequencies, which push the quantum regime to higher temperatures (e.g. $1\,\mathrm{THz}\approx 48\,\mathrm{K}$), the relatively short lifetimes currently limit the use of hematite, $\mathrm{YFeO_3}$ and other antiferromagnets for quantum magnonics.
 
\subsection*{Altermagnets}
Recently, altermagnets have emerged as a distinct class of magnetic materials beyond conventional ferro- and antiferromagnets (see Fig.\,\ref{f:1}\,(c)), characterized by compensated collinear order in real space and non-relativistic alternating spin splitting in momentum space \cite{song2025altermagnets, gomonay2024structure}. This unique duality yields terahertz-frequency, field-robust dynamics with negligible stray fields—features reminiscent of antiferromagnets\,—\,while also enabling spin-polarized band structures and novel transport phenomena, such as the crystal anomalous Hall effect, the spin-splitter effect, and anisotropic magnon spectra. Depending on the orbital symmetry of the momentum-dependent spin splitting, altermagnets are classified into even-parity (d-, g- and i-wave) orders with nodal sign changes (e.g., d-wave materials \cite{song2025altermagnets, zarzuela2025transport} like RuO$_2$ \cite{gomonay2024structure}, MnTe\cite{amin2024nanoscale}). They are distinguished from s‑ and p‑wave (odd or no‑node) patterns found in ferromagnets or antiferromagnets. A prominent example is hematite ($\alpha$-Fe$_2$O$_3$), a well-established antiferromagnet in magnonics\cite{Chen2025,Hamdi2023,HWang2023,ElKanj2023,Gckelhorn2023}, which has recently been identified as an altermagnet and shown to host anisotropic magnon excitations consistent with momentum-dependent spin splitting\cite{Hoyer2025,GalindezRuales2025}. Altogether, this makes altermagnets attractive for quantum magnonics, although the reduced lifetimes due to magnon–magnon interactions\cite{GarciaGaitan2025} should be taken into account, as is the case for antiferromagnets.

\subsection*{2D van der Waals and organic magnets}
Recently, two-dimensional (2D) van der Waals magnets, illustrated in Fig.\,\ref{f:1}\,(d) and (e), have attracted renewed interest in magnonics\cite{Kruglyak2024}. A notable example is the van der Waals antiferromagnet CrPS$_4$ (see Tab.\,\ref{tab:materials} and Fig.\,\ref{f:1}\,(e)), which supports long-distance magnon transport\cite{deWal2023}. Likewise, van der Waals magnets, such as CrCl$_3$ exhibit standing spin waves. Experiments resolve modes across a thickness of $20\,$µ$\mathrm{m}$, demonstrating control of spin-wave modes in 2D systems\cite{Kapoor2020}. On the theoretical side, theory shows that dipolar interactions strongly shape magnon dispersion and edge-localized modes in two-dimensional van der Waals ferromagnets\cite{Hussain2022}. Propagating spin waves have been detected in the 2D van der Waals ferromagnet Fe$_5$GeTe$_2$, despite relatively high magnetic damping\cite{Schulz2023}. In addition, recent advances in materials engineering have demonstrated that interfacial control can substantially enhance the magnetic properties of van der Waals magnets, as exemplified by wafer-scale Fe$_4$GeTe$_2$ films exhibiting robust ferromagnetism far above room temperature, emphasizing the potential of interface engineering to tailor magnetic order and magnonic properties in two-dimensional systems\cite{HaWang2023}. These results highlight 2D materials as tunable and versatile platforms for future magnonic applications. 

Organic magnets also offer promise through chemical tunability and potentially low-damping propagation. For example, V(TCNE)$_x$ exhibits lifetimes up to 40\,ns\cite{Liu2020,McCullian2020}. Low magnetic damping, together with their other advantageous properties, makes these materials highly promising for quantum magnonics. Further research is needed to fully understand and explore their potential.
\subsection*{Hexaferrites}

For long magnon lifetimes and extended free paths, it is essential to use electrically insulating materials to eliminate the dissipation channel from magnons into free electrons. A prominent class of such magnetic insulators are the M-type hexaferrites, with barium hexaferrite (BaFe$_{12}$O$_{19}$, BaM) as their flagship representative. BaM is an insulating ferrimagnet with robust performance across a wide temperature range ($T_\mathrm{C} \approx 725$~K)~\cite{Harris2012}, high coercivity and remanence, and a saturation magnetization of $M_\mathrm{s} \approx 380$~kA/m at RT~\cite{Harris2012}. Its strong intrinsic magnetic anisotropy ($H_\mathrm{a} \approx 1353$~kA/m) enables broadband frequency operation (50--65~GHz in undoped films at low magnetic fields~\cite{song2010self}, and 20--100~GHz in doped films~\cite{Levchenko2024} under moderate biasing magnetic fields $<0.8$~T), as well as comparatively low magnetic losses (Gilbert damping constants as low as $\alpha \sim 7\times10^{-4}$)~\cite{li2016spin,malkinski2012advanced}. 

BaM supports spin waves with high group velocities, while its strong perpendicular magnetic anisotropy permits isotropic forward-volume spin-wave propagation in the film plane already under low external magnetic bias. Combined with intrinsically low magnetic damping, with Gilbert damping constants on the order of $\alpha \sim 10^{-4}$, these properties make BaM a highly attractive material for magnonic radio-frequency (RF) signal processing and data-processing applications, including operation in the 5G FR2 frequency range (24.25--71.0\,GHz)~\cite{5GNR,Levchenko2024}. BaM uniquely enables magnonic access to technologically relevant frequency bands above 60\,GHz that are difficult to reach with other ferrimagnets. These include the 60\,GHz band used for in-cabin automotive radar and health-monitoring applications~\cite{Infineon60GHzRadar}, as well as the 76--81\,GHz band employed in long-range automotive radar systems~\cite{IBLenhardtAutomotiveRadar}. Importantly, the large uniaxial anisotropy enables self-magnetized operation, allowing access to high-frequency magnonic modes with little or no external bias field. In nm-thick BaM waveguides, the fundamental spin-wave frequencies naturally lie in the tens-of-GHz range. These frequencies are primarily set by the internal anisotropy field rather than an applied magnetic field.

By providing high-frequency and potentially field-free spin-wave operation, BaM positions magnonics as a viable and competitive platform for next-generation RF signal processing and sensing technologies, enabling contributions to application domains that remain inaccessible to conventional magnonic materials. At the same time, the intrinsically high operating frequencies and low damping of BaM make this material class a promising candidate for high-frequency quantum magnonics, where large magnon energies are advantageous for suppressing thermal occupation in quantum magnonics. The main limitation of BaM for modern magnonics arises from the absence of lattice-matched growth techniques capable of producing high-quality single-crystal nanoscale films that simultaneously provide low magnetic loss, low bias fields, and high-frequency performance. Systematic studies of BaM at cryogenic temperatures are therefore highly desirable and remain largely unexplored.
\vspace{-1em}
\subsection*{Europium Chalcogenides}
A class of magnetic insulators and semiconductors that become magnetically ordered only at cryogenic temperatures\,---\,and are therefore unsuitable for room-temperature magnonics and spintronics, yet potentially relevant for quantum magnonics\,---\,are the europium chalcogenides (Eu$^{2+}$X$^{2-}$)\cite{Wachter1979,Schlipf2013,Boncher2015,Dietl2019}. This material family includes europium oxide (EuO), europium sulfide (EuS), europium selenide (EuSe), and europium telluride (EuTe). All compounds crystallize in the rock-salt structure and exhibit band gaps ranging from approximately 1.2\,eV (EuSe) to about 2.4\,eV (EuTe), making them model systems for studying localized Heisenberg magnetism driven by 4f electrons\cite{Kolokolov2025}.

The magnetic order in europium chalcogenides is governed by nearest- and next-nearest-neighbor exchange interactions, resulting in ferromagnetic ground states in EuO and EuS with Curie temperatures of $T_\mathrm{C}\approx69\,\mathrm{K}$ and $T_\mathrm{C}\approx16.6\,\mathrm{K}$, respectively. In contrast, EuSe exhibits a complex antiferromagnetic ground state below $T_\mathrm{N}\approx4.6\,\mathrm{K}$ with field-induced metamagnetic phases that depend sensitively on temperature and external magnetic field, while EuTe is an antiferromagnet with a Néel temperature of $T_\mathrm{N}\approx9.6\,\mathrm{K}$.
Owing to the large magnetic moment of the Eu$^{2+}$ ion ($S=7/2$), these materials exhibit exceptionally high saturation magnetizations, reaching approximately 1900\,kA/m in EuO and about 1300\,kA/m in EuTe (in the fully field-saturated antiferromagnetic state). As a result, spin-wave frequencies in europium chalcogenides are significantly higher than in most conventional magnetic insulators.

Magnetic damping and resonance linewidths in europium chalcogenides have been investigated primarily at low temperatures using ferromagnetic and antiferromagnetic resonance techniques. In high-quality single crystals, remarkably narrow linewidths have been reported, with values on the order of millitesla for EuS, while comparable or slightly larger low-temperature linewidths have been observed in EuO, EuSe, and EuTe depending on the magnetic phase and field configuration\cite{vonMolnar1965,Franzblau1967,Brown1968,Battles1969,Eastman1968,Kunii1974,Dillon1964}. Taking the lowest reported linewidth for EuS, $\Delta B \approx 1\,\mathrm{mT}$ at $T\approx1.4\,\mathrm{K}$ and microwave frequencies around 22\,GHz, yields a magnon lifetime of approximately $\tau \approx 6\,\mathrm{ns}$\cite{Franzblau1967}.

EuS has recently been employed in studies of magnon transport driven by the spin Seebeck effect in ferromagnetic EuS thin films\cite{AguilarPujol2023}, demonstrating its relevance for cryogenic magnonics. While the magnon lifetimes in europium chalcogenides are currently much shorter than those achieved in quantum magnonics experiments based on YIG, they are already comparable to coherence times relevant for hybrid quantum systems operating in the few-gigahertz regime. Given the localized 4f magnetism, the absence of itinerant charge carriers, and the presence of strong exchange fields, further improvements in linewidth and magnon lifetime may be expected in defect-reduced samples and at millikelvin temperatures. These considerations position europium chalcogenides as a promising, yet largely unexplored, materials platform for quantum magnonics and motivate systematic investigations of their magnon lifetimes, coherence properties, and dissipation mechanisms in the quantum limit.
\vspace{2em}

\subsection*{Yttrium Iron Garnet}
The longest magnon lifetimes have so far been achieved in YIG, which remains the only material particularly used for single-magnon excitation, measurement, and manipulation. This outstanding material is a ferrimagnetic insulator\cite{SagaYIG,YIGmagnonics} has a well understood magnetization behavior\cite{Serha2025c} and magnetic anisotropy\cite{Bobkov1997}. The damping of coherent magnons arises primarily from scattering on crystallographic and surface defects\cite{Spencer1959,Melkov2004}, from rare-earth impurities\cite{Serha2025b,Michalceanu2018,PardaviHorvath2000,Dillon1959,Spencer1961}, and from interactions with thermal magnons\cite{Pincus1961,White1963} (spin-spin interaction) and phonons\cite{Kasuya1961,Klein2004,Naletov2007,Rckriegel2014,Streib2019,Mller2024,Ba2025,Cherkasskii2025} (spin-lattice interaction). As summarized in Tab.\,\ref{tab:materialsfilm}, for example at millikelvin temperatures, an FMR linewidth of $0.02\,\mathrm{mT}$ corresponds to a magnon lifetime of about $300\,\mathrm{ns}$\cite{Serha2025b}, i.e., the minimally required timescale of a few hundred nanoseconds for single-magnon experiments\cite{Lachance-Quirion2020,Xu2023}. Even so, lifetimes of this order are still sufficient to sustain spin-wave coherence for more than 10\,µs \cite{Makiuchi2024} and to support magnon propagation over distances exceeding 100\,µm using a 110\,nm-thin YIG liquid-phase epitaxy (LPE) film.\cite{Bensmann2025}.

For advanced quantum magnonic operations, longer magnon lifetimes are required. Previous experiments used the Kittel mode, a uniform precession of the magnetization across the whole sample. The damping of the FMR mode and dipolar magnons with large wavelengths is dominated by two-magnon scattering from lattice and surface defects, and is therefore largely temperature independent. On cooling to millikelvin temperatures, the lifetime decreases only slightly compared with RT\cite{Kosen2019,Schmoll2024,Serha2025b}. In contrast, short-wavelength magnons with finite wave vector, that are localized inside the sample, are insensitive to surface quality. For these dipolar-exchange spin waves (DESW) the dominant dissipation is scattering from thermal magnons and phonons. Recently, Serha\textit{ et al.}\cite{Serha2025b} showed that when this thermal bath is depleted the lifetime of short-wavelength magnons increases by more than an order of magnitude and reaches up to 18\,µs. These magnons were generated through the intrinsic nonlinearity of the system via three-magnon splitting, with the lifetime set by the power threshold for entering the nonlinear regime\cite{Suhl1957}. 

The lifetime of these secondary magnons increases drastically as the temperature decreases from a few kelvin to the millikelvin range, as shown in Fig.\,\ref{f:1}\,(f). As the temperature approaches absolute zero, multimagnon scattering on thermal magnons and interactions with thermal phonons vanish, and because short-wavelength DESW remain localized within the crystal volume and are effectively decoupled from dipolar coupling to surrounding dissipation channels, their intrinsic relaxation is strongly suppressed, resulting in a pronounced enhancement of the magnon lifetime at millikelvin temperatures\cite{Serha2025b}.

In theory, the lifetime of these magnons is expected to approach infinity at millikelvin temperatures. In practice, however, it saturates below about 250\,mK\cite{Serha2025b}. Interestingly, this saturation occurs at different lifetime levels, with crystal contamination with rare and heavy metal ions becoming the limiting factor. This impurity-induced lifetime-limiting bottleneck is evident in Fig.\,\ref{f:1}\,(f), where a reduction in impurity concentration from Sphere~1 to Sphere~3 leads to an increase in the saturation lifetime of DESW magnons from approximately 5\,µs to 18\,µs. As higher-purity samples exhibit longer magnon lifetimes, indicating that values beyond 20\,µs\,---\,the dephasing time $T_2$ of state-of-the-art superconducting qubits \cite{Burnett2019}\,---\,are within reach when YIG is grown from specially purified yttrium oxide ($\mathrm{Y_2O_3}$), which contains most of the rare-earth and $L$-state transition metal ion responsible for limiting YIG quality at low temperatures.

In total, DESW exhibit lifetimes at millikelvin temperatures that are long enough to transport quantum information and entanglement in micrometer- and submicrometer-thick films, for example between superconducting qubits. This paves the way for new experiments and applications in quantum magnonics, where magnons act as efficient carriers of quantum information.

\section{\label{sec:Films} YIG films for quantum magnonics}
\begin{table*}
\caption{\label{tab:materialsfilm}Comparison of bulk YIG and thin YIG films relevant to quantum magnonics.}
\begingroup
\renewcommand{\arraystretch}{1.25}
\setlength{\tabcolsep}{6pt}
\begin{ruledtabular}
\begin{tabular}{p{0.4\textwidth} p{0.1\textwidth} 
                p{0.1\textwidth} p{0.1\textwidth} p{0.1\textwidth}}
\textbf{Property} &
\pcell{0.12\textwidth}{bulk YIG\cite{Serha2025b,Maier-Flaig2017,Klingler2017}} &

\pcell{0.11\textwidth}{YIG/GGG\cite{Serha2025d}} &
\pcell{0.11\textwidth}{YIG/YSGG\cite{Youssef20205}} &
\pcell{0.11\textwidth}{YIG/YSGAG\cite{Serha2025d}} \\
\hline
Saturation magnetization $M_{\mathrm{_S}}$ @ RT\,\,|\,\,$\rightarrow$\,0\,K (kA/m) &
\pcell{0.11\textwidth}{140\,\,|\,\,200} &

\pcell{0.11\textwidth}{144\,\,|\,\,205} &
\pcell{0.11\textwidth}{67\,\,|\,\,95} &
\pcell{0.11\textwidth}{131\,\,|\,\,184} \\

Gilbert damping $\alpha$ @ RT ($\times10^{-5}$)&
\pcell{0.11\textwidth}{$3\pm0.5$} &

\pcell{0.11\textwidth}{$4.3\pm0.4$} &
\pcell{0.11\textwidth}{$10\pm1$} &
\pcell{0.11\textwidth}{$4.3\pm0.6$} \\

Inhomogeneous linewidth broadening $\Delta B_0$ @ RT (mT) &
\pcell{0.11\textwidth}{$0-0.05$} &

\pcell{0.11\textwidth}{$0.14$} &
\pcell{0.11\textwidth}{$0.50$} &
\pcell{0.11\textwidth}{$0.15$} \\

FMR linewidth $\Delta B\,@\approx8$\,GHz @ RT\,\,|\,\,$\rightarrow$\,0\,K (mT)  &

\pcell{0.11\textwidth}{$0.03\,\,|\,\,0.02$} &

\pcell{0.11\textwidth}{$0.19\,\,|\,\,0.85$} &
\pcell{0.11\textwidth}{$0.50\,\,|\,\,0.75$} &
\pcell{0.11\textwidth}{$0.17\,\,|\,\,0.25$} \\


\end{tabular}
\end{ruledtabular}
\endgroup
\end{table*}
\begin{figure}
\includegraphics{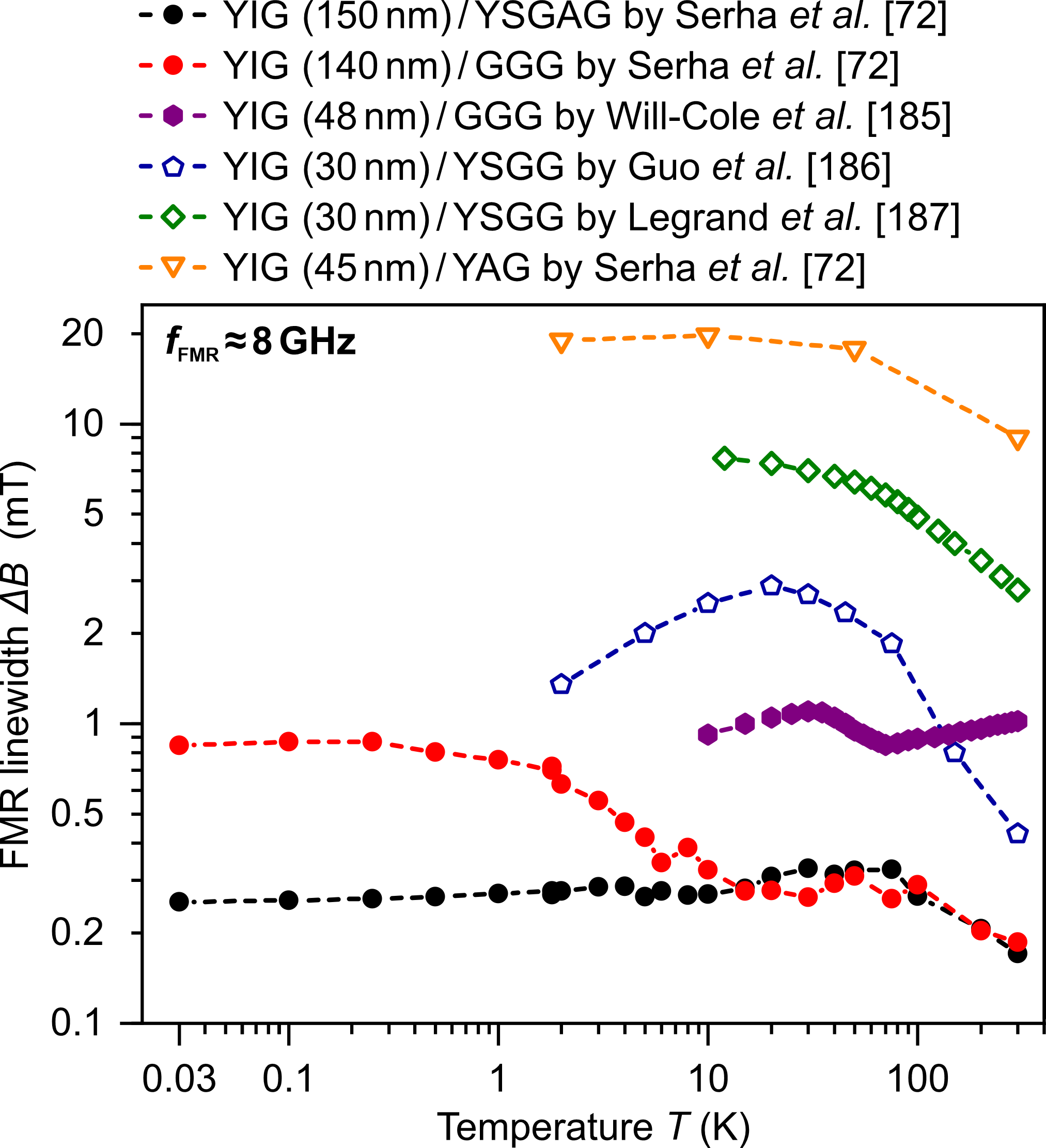}
\caption{\label{f:2} FMR FWHM linewidth $\Delta B$ (measure for magnetic damping) as a function of temperature $T$ for an FMR frequency $f_{\mathrm{FMR}} \approx 8\,\mathrm{GHz}$, displayed on a double logarithmic scale. The graph compares measurements obtained in this work with selected results of sputtered (hollow symbols) and LPE grown (filled symbols) YIG films from relevant literature, enabling a direct evaluation of temperature-dependent damping behavior across different samples, their substrates and studies \cite{Cole2023,Guo2023,Legrand2024,Serha2025d}. Reproduced from R.~O.~Serha \textit{et al.},  arXiv:2508.19044 (2025); licensed under a Creative Commons Attribution-NonCommercial-NoDerivatives 4.0 International (CC BY-NC-ND) license.}
\end{figure}

To use magnons as quantum data carriers in quantum magnonics, it is crucial to move beyond standing magnons in spherical resonators\cite{Lachance-Quirion2019,Lachance-Quirion2020,Xu2023,Li2022,Rani2025} and establish YIG films as the active medium for propagating single magnons. At RT, the highest bulk-like quality YIG films, from micrometer to nanometer scales, have been grown on paramagnetic GGG substrates because of the nearly perfect lattice matching between the two crystals \cite{Dubs2017,Dubs2020,Ding2020,Heyroth2019,Cornelissen2015,Onbasli2014,Hahn2014,Askarzadeh2025}.

However, quantum magnonics must operate at very low temperatures down to the millikelvin range. Under these conditions, paramagnetic substrates, such as GGG become readily magnetized by an externally applied magnetic field \cite{Danilov1989,Serha2024}. Counterintuitively, the GGG substrate does not behave as a simple paramagnet describable by a Brillouin function at temperatures below 500\,mK. Instead, in this temperature range it exhibits a complex magnetic phase diagram and enters a frustrated magnetic state, in which the magnetization becomes nearly temperature independent and is governed primarily by the external magnetic field\cite{Serha2024,Deen2015,Petrenko1998,Schiffer1994}. The magnetized substrate generates an inhomogeneous stray field that depends on its geometry and imposes a strong gradient in the internal field of the YIG film. For example, a common $(5\times5\times0.5)$\,mm$^3$ GGG substrate at 1.8\,K and an in-plane bias field of 600\,mT produces stray fields of up to 60\,mT at the edges and about 12\,mT at the center opposing the external field\cite{Serha2024}. This gradient is already broadening the FMR linewidths measured over the sample\cite{Serha2025}. By selecting an appropriate substrate geometry or by microstructuring the YIG film, the GGG-induced stray-field gradient can be eliminated across the magnetic medium, resulting in a homogeneous internal field in YIG\cite{Schmoll2025}. However, the partially magnetically ordered substrate couples to the YIG spin system \cite{Schmoll2024}, opening extra dissipation channels that broaden the ferromagnetic resonance (FMR) linewidth \cite{Serha2025,Serha2024,Guo2022,Jermain2017,Michalceanu2018,Kosen2019} and reduce magnon lifetimes of propagating dipolar magnons \cite{Schmoll2024}. The linewidth broadening effect is shown in Fig.\,\ref{f:1}\,(g)\cite{Serha2025} and Tab.\,\ref{tab:materialsfilm} reports a linewidth difference of more than a factor of four between RT and $T\,\rightarrow 0$, even in the absence of the inhomogeneous GGG-induced stray field. Such a decrease in magnon lifetime undermines the advantages that YIG would otherwise offer for quantum-magnonic experiments. Microstructuring\cite{Schmoll2025} and operating at low external fields\cite{Knauer2023} can mitigate the stray-field gradient to some extent. Nevertheless, eliminating the coupling between the substrate spin system and the YIG film requires the use of diamagnetic substrates. For a comparison of the temperature-dependent damping behavior between YIG films grown on GGG and on diamagnetic alternative substrates, Fig.\,\ref{f:2} presents literature results on sputtered and LPE-grown YIG films from different studies\cite{Cole2023,Guo2023,Legrand2024,Serha2025d}.

Yttrium aluminum garnet (YAG) was among the first diamagnetic substrates proposed to replace paramagnetic GGG in YIG heterostructures, aiming to unlock new possibilities for quantum magnonics at cryogenic temperatures by avoiding substrate magnetization in applied fields. However, despite the drawbacks of GGG at low temperatures, YIG/GGG still outperforms YIG/YAG at both RT and low temperature (see Fig.\,\ref{f:2}), primarily because the large lattice mismatch between YIG and YAG degrades crystalline quality of the YIG film\cite{Sposito2013,Krysztofik2021,Serha2025d}.

In general, the broad garnet family \cite{INBOOKNomura1978,Yoshimoto2019} offers promising diamagnetic substrates with lattice constants closely matching YIG. For example, yttrium scandium aluminum garnet (YSAG), used as a diamagnetic spacer between YIG and GGG, has been shown to reduce damping in YIG films at low temperatures \cite{Guo2022}. A stronger candidate is yttrium scandium gallium garnet (YSGG), which can serve directly as a substrate\cite{Guo2023,Legrand2024}. YIG/YSGG has caught up with YIG/GGG in terms of FMR linewidth at low temperatures\cite{Youssef20205} (see Tab.\,\ref{tab:materialsfilm}), and even demonstrated robust and superior spin-wave propagation at 2\,K\cite{Abrao2025}. However, this improvement comes at the cost of lower saturation magnetization, which yields slower dipolar magnons, as the group velocity of magetostatical spin waves are pratically directly proportional to the magnetization saturation. In addition, for YIG/YSGG, only thin films of about 60\,nm can be grown fully strained due to the large lattice mismatch, and these still do not reach the best room-temperature damping values achieved by YIG grown on GGG substrates\cite{Serha2025d}.

A breakthrough in the field came with the introduction of yttrium scandium gallium aluminum garnet (YSGAG) by C.~Guguschev and C.~Dubs \textit{et al.}~\cite{Dubs2025}. This diamagnetic garnet enables improved lattice matching up to no mismatch with YIG. Figure\,\ref{f:1}\,(h) illustrates the $\mathrm{Y}_3\mathrm{(Sc,Ga,Al)}_2\mathrm{(Ga,Al)}_3\mathrm{O}_{12}$ crystal structure, highlighting selected dodecahedral, octahedral, and tetrahedral oxygen coordination polyhedra with the corresponding ions at their centers. By customizing the distribution ratio of the diamagnetic ions Sc, Ga, and Al on the octahedral and tetrahedral sites, the lattice parameter can be tuned to match that of epitaxial YIG films~\cite{Dubs2025}. As shown in Fig.\,\ref{f:2}, taken from the study\cite{Serha2025d}, YIG films grown by LPE on YSGAG exhibit simultaneously the same low damping at RT and at millikelvin temperatures that YIG on GGG achieves only at RT, thereby eliminating the drawbacks at cryogenic temperatures associated with paramagnetic substrates. Using YSGAG as the substrate, record-low FMR damping in YIG films were observed also at RT, with FMR linewidths as narrow as $0.1\,\mathrm{mT}$ at frequencies up to $4\,\mathrm{GHz}$ and a Gilbert damping parameter $\alpha = 4.29\times10^{-5}$.

From the properties summarized in Tab.\,\ref{tab:materialsfilm}, it is clear that the YIG/YSGAG system comes closest to bulk YIG at cryogenic temperatures. Its magnon lifetime is still shorter by a factor of about 5–10 compared with ultra-pure bulk YIG \cite{Serha2025b}. However, YIG LPE film quality is known to improve with increasing thicknesses\cite{Dubs2017}, and careful co-tuning of the YIG and YSGAG lattice parameters can yield an essentially perfect match of both systems. This approach enables YIG films with bulk-like magnetic quality at millikelvin temperatures—an ideal platform for studying single-magnon propagation in quantum magnonics.

\section{\label{sec:Out} Outlook}
Quantum magnonics is inherently materials-driven, with ongoing efforts focused on identifying magnetic systems whose properties best match the demands of coherence, frequency, coupling strength, and scalability for a given application. Ferromagnetic metals and Heusler compounds offer large saturation magnetization, strong exchange, and excellent compatibility with nanofabrication, enabling strong coupling to microwave photons, spin currents, and superconducting circuits, but their finite electrical conductivity introduces intrinsic dissipation that presently limits magnon lifetimes to the nanosecond regime. Organic magnets provide chemically and structurally tunable systems and have demonstrated comparatively long lifetimes for their class, reaching tens of nanoseconds, yet remain constrained by disorder, interfacial effects, and the lack of systematic investigations at millikelvin temperatures. Magnetic insulators with high magnetization saturation or anisotropy, such as hexaferrites and europium chalcogenides, open access to high-frequency magnon modes and large magnon energies, offering prospects for bias-field-free operation in waveguides and high-frequency quantum magnonics, although challenges in thin-film growth, and low-temperature damping must still be addressed. Taken together, these diverse material classes highlight that future progress in quantum magnonics will rely on targeted materials optimization across multiple platforms rather than on a single universal material system.

Within this diverse materials landscape, two converging breakthrough are leading to the perspective using ultra-long-living propagating magnons for future solid-state quantum technologies. First, diamagnetic YSGAG substrates with a compositionally tunable lattice can be matched to YIG for nanometer-thin as well as micrometer-thick films. This approach removes substrate magnetization and stray-field gradients at cryogenic temperatures, while preserving the room-temperature damping performance that has made YIG/GGG films the benchmark. Second, in ultra pure YIG up to 18\,µs microsecond lifetimes for short wavelength dipolar–exchange magnons were demonstrated once the thermal magnon bath is depleted, showing that long coherence is available when impurities and extrinsic scattering are suppressed. Together, these results point to YIG films that support single magnon excitation, their propagation, and manipulation on-chip opening new directions for solid-state quantum magnonics.

A practical route to this goal is already provided by nanometer-thin YIG films grown on YSGAG. For submicrometer- or micrometer-thick films, an even better lattice match between the substrate and YIG is required. Such a perfectly matched, diamagnetic system would avoid the cryogenic drawbacks of paramagnetic GGG or strained substrate/YIG combinations, allowing the film to approach the intrinsic low dissipation of bulk YIG. In this regime, magnon lifetimes are governed mainly by material imperfections, making growth and processing critical: impurity levels must approach ultra-pure single-crystal quality with strict control of rare-earth contamination, while careful surface and edge preparation suppresses two-magnon scattering. At millikelvin temperatures, dipolar-exchange magnons reach lifetimes above ten microseconds. A remaining obstacle is that DESWs achieve long lifetimes by being insensitive to surfaces, but this also implies the low group velocity of dipolar waves, and their lifetime is expected to decrease with the increase in velocity. One potential solution would be to use low-frequency exchange waves with wavelengths shorter than the DESW, with reduced two-magnon scattering efficiencies and high group velocities. 

Taken together, these elements will provide access to long-living and long-propagating single magnons that can be used as magnonic quantum information carriers. These carriers can serve as entanglement busses or gates between superconducting qubits placed on the same YSGAG substrates. Consequently, such single magnons will enable the integration of 'classical magnonics' data processing approaches, including RF applications, Boolean and neuromorphic computing, with magnon-based quantum computing concepts on the same nanoscale chip.


\begin{acknowledgments}
This research was funded in part by the Austrian Science Fund (FWF) project Paramagnonics [10.55776/I6568].
The work was supported by the German Federal Ministry of Research, Technology and Space (BMFTR) under the reference numbers (13N17109) within a collaborative project “Low-loss materials for integrated magnonic-superconducting quantum technologies (MagSQuant)”.
The authors thank Khrystyna O. Levchenko for insightful consultations and valuable scientific discussions on magnetic materials, in particular on hexaferrites.
\end{acknowledgments}

\section*{Data Availability Statement}
Data sharing is not applicable to this article as no new data were created or analyzed in this study.

\bibliography{BIB}

\end{document}